\documentclass[aps,prc,floatfix,showpacs,twocolumn]{revtex4-2}
\usepackage{graphicx}
\usepackage[breaklinks,colorlinks,urlcolor=blue,citecolor=blue]{hyperref}
\usepackage{soul}
\usepackage{color}
\usepackage{xcolor}
\usepackage[normalem]{ulem}
\usepackage[utf8]{inputenc}
\usepackage{graphics}
\usepackage{amsmath}
\usepackage{ amssymb }
\usepackage[export]{adjustbox}

\usepackage[utf8]{inputenc}

\begin{document}

\title{Decision Theory for the Mass Measurements at the
  Facility for Rare Isotope Beams}
\author{Jesse N. Farr$^{1}$}
\author{Zach Meisel$^{2}$}
\author{Andrew W. Steiner$^{1}$}
\affiliation{$^{1}$Department of Physics and Astronomy, University of
  Tennessee, Knoxville, TN 37996, USA}
\affiliation{$^{2}$Institute of Nuclear \& Particle Physics, Department of Physics \& Astronomy, Ohio University, Athens, OH, 45701 USA}

\begin{abstract}

Nuclear physics facilities, like the Facility for Rare Isotope Beams
(FRIB), can potentially perform many nuclear mass measurements of
exotic isotopes. Each measurement comes with a particular cost, both
in time and money, and thus it is important to establish which mass
measurements are the most informative. In this article, we show that
one can use the Kullback-Leibler divergence to determine the
information gained by a mass measurement. We model the information
gain obtained by nuclear mass measurements from two perspectives:
first from the perspective of theoretical nuclear mass models, and the
second from the perspective of r-process nucleosynthesis. While this
work specifically analyzes the abilities of FRIB, other facilities
worldwide could benefit from a similar use of information gain in
order to decide which experiments are optimal.
\end{abstract}

\maketitle

\section{Introduction}

There are thousands of isotopes accessible using nuclear physics
facilities like the Facility for Rare Isotope Beams (FRIB), yet each 
experiment requires beam time and a considerable amount of human
effort. Thus, it is worth attempting to quantify the information
gained by performing an experiment relative to its cost. One mechanism
for making that quantification is decision theory, where a utility
function is maximized in order to make the best decision given a
domain of problems under consideration. The utility function, however,
is not uniquely determined. It depends on the nature of the facility,
the results which have been obtained from previous facilities, and
even the makeup of the team performing the next experiment. 

 We choose only to analyze nuclear mass measurements at FRIB, and
 leave the consideration of other facilities to future work. We also
 make the (strong) assumption that this cost is dominated by the beam
 time required to perform a mass measurement.

Given a prior probability distribution $P(x_1,x_2,\ldots,x_k)$ defined
over a domain $X$, and a resulting posterior distribution $Q$, the
information contained in the posterior relative to the prior is the
Kullback-Leibler (KL) divergence
\begin{equation}
D_{KL}(Q||P)=\sum_{x\in{X}}P(x)\ln\frac{P(x)}{Q(x)} \, .
\label{eq:kldivgaus}
\end{equation}
If the posterior and prior are identical, then the KL divergence is
zero. Under the additional assumption that both the prior and
posterior distributions are multivariate Gaussians, the KL divergence
is
\begin{eqnarray}
  D_{\mathrm{KL}}({\cal N}_Q||{\cal N}_P)&=&\frac{1}{2}\left[\mathrm{tr}
    (\Sigma_P^{-1}\Sigma_Q)-k+\ln\left(\frac{\mathrm{det}~\Sigma_P}
    {\mathrm{det}~\Sigma_Q}\right) \right. \nonumber \\
&& + \left. (\mu_P-\mu_Q)^T\Sigma_P^{-1}(\mu_P-\mu_Q)\right]
\label{eq:kldiv}
\end{eqnarray}
where ${\cal N}_P$ and ${\cal N}_Q$ are the prior and posterior
distributions, respectively, $\Sigma_P$ and $\Sigma_Q$ are the
corresponding covariance matrices, $\mu_P$ and $\mu_Q$ are the means
of the Gaussians, and $k$ is the dimensionality of the space. In the
case of two one-dimensional Gaussians with identical means, the KL
divergence is simply a function of the two standard deviations, i.e.
$D_{\mathrm{KL}}(\sigma_Q | \sigma_P)$ and the KL divergence depends
only on the ratio of the two, $\sigma_P/\sigma_Q$.

The value of nuclear mass measurements, however, lies not only in the
information contained in the masses themselves, but also in what those
masses mean for theoretical models of nuclear structure or physical
processes involving nuclei. Both of these applications come with their
own information gain. In this work, we model the information gain
obtained by nuclear mass measurements from two perspectives: first
from the perspective of theoretical mass models as understood by mass
tabulations, and the second from the perspective of r-process
nucleosynthesis.

\section{Estimated Mass Reach}

The two primary mass measurement techniques to be employed at FRIB are
Penning trap mass spectrometry (PTMS) and time-of-flight mass
spectrometry (TOFMS). PTMS is a high precision technique requiring
stopped radioactive ion beams, while TOFMS is a lower precision
technique that uses fast beams and is suited for nuclides with short
half-lives or low production rates. Generally speaking, PTMS is used
to establish a precise nuclear mass surface that TOFMS then extends to
more exotic isotopes.

PTMS measurements at FRIB will be performed with the Low Energy Beam
Ion Trap~\cite{Reds13}. Several PTMS techniques exist. Here we
consider the time-of-flight ion-cyclotron-resonance (TOF-ICR)
technique, which is the most commonly used to date. For the TOF-ICR
PTMS technique, the ion's cyclotron resonance is found and converted
from the orbital motion into motion leaving the trap, which is
proportional to the ion's mass. The relative statistical uncertainty
$\delta m/m$ is roughly given by $\delta m/m\approx R^{-1}n^{-1/2}$,
where $n$ is the number of ions detected and $R$ is the resolving
power~\cite{Lunn03}. The resolving power is approximately equal to the
product of the cyclotron frequency of the ion in the trap (typically
$\sim$~MHz) and the length of time the ion orbits in the trap
$t_{\rm{obs}}$ (typically $\sim$0.1~s). $R$ therefore depends on many
considerations, such as the mass of the nucleus of interest, the
obtainable charge-state, the time it takes to produce the optimum
charge state, and the nuclear half-life. Given the uncertainties in
charge-breeding capabilities and the approximate nature of the
estimate for $n$, we assume $R=10^{5}$, which is in-line with sample
cases for rare isotopes~\cite{Boll01}. Further, we assume a typical
$t_{\rm{obs}}$=100~ms, meaning that $n$ is the product of the FRIB
stopped-beam rate and the duration of the experiment, assumed to be 24
hours, reduced by the radioactive decay of ions during the measurement
process. Experimental $\beta$-decay half-lives~\cite{Tuli11} are used
when available and theory estimates~\cite{Moll03} are used otherwise.
The assumed systematic uncertainty is $\delta
m/m|_{\rm{syst}}=10^{-8}$, which is typical for the measurement
precision of reference ions used in a PTMS measurement~\cite{Reds13}.

TOFMS at FRIB will consist of TOF and magnetic rigidity measurements
for ions traversing a several tens of meter flight path, e.g. as in
Ref.~\cite{Meis20}. The proportionality between rigidity-corrected TOF
and nuclear mass is calibrated by simultaneously measuring nuclides
with known masses, typically from prior PTMS measurements. The
statistical uncertainty of TOFMS is related to the TOF measurement
precision $\sigma_{\rm{TOF}}$/TOF and $n$, here based on the fast beam
rate and 100 hours of measurement time, by $\delta
m/m\approx\sigma_{\rm{TOF}}/({\rm TOF}\sqrt{n})$, where we assume a
typical $\sigma_{\rm{TOF}}/\rm{TOF}$ of $10^{-4}$~\cite{Meis20b}. The
systematic uncertainty that we adopt is based on the empirically
motivated approximation that $\delta
m/m|_{\rm{syst}}=5\times10^{-6}(1+(N-N_{\rm{ref}}))$, where
$N-N_{\rm{ref}}$ is the number of neutrons separating the nuclide of
interest and the most neutron-rich isotope of that element with a
known mass uncertainty $\leq10^{-6}$~\cite{Meis20b}.

Using the estimated fast and stopped beam rates at FRIB~\cite{Boll11},
we calculate $n$. We then estimate $\delta m/m$ for PTMS and, for
cases where this is greater than $10^{-6}$, for TOFMS, up to cases
with $\delta m/m=10^{-4}$. The smaller of the two $\delta m$ is used
in the subsequent calculation of the KL divergence.

\section{Information Gain Relative to Theoretical Mass Models}

For each nucleus which from the previous section which is accessible
from FRIB, we presume that the associated probability distribution
from theory is a Gaussian, given by the mean and the standard
deviation of the theoretical predictions across mass models. Thus
\begin{equation}
    \sigma_{\mathrm{th}}=\sigma_{M(Z,N)} \, .
\end{equation} 
For the theoretical mass models, we use the masses from
%we include the 2020 Atomic Mass Evaluation~\cite{Huang21ta}, 
Chamel et al. (2009)~\cite{HFB17}, Duflo et al. (2003)~\cite{DZ},
Ebran et al. (2011)~\cite{HFB24}, Goriely et al. (2003, 2007, 2008,
2008b, 2010, 2014)~\cite{HFB2,HFB8,HFB14,HFB21,HFB25,HFB27}, Koura et
al. (2005)~\cite{Ktuy2005}, Long et al. (2010)~\cite{HFB22}, Liu et
al. (2011)~\cite{Liu11}, Moller et al. (2016)~\cite{Moller16ng},
Pearson et al. (2011)~\cite{HFB26}, Rath et al. (2010)~\cite{HFB23},
and Wang et al. (2010, 2001b) ~\cite{Wang10,Wang10b}. The decision to
exclude other similar mass models was made in part due to the limited
range of isotopes that the models cover. We have found that our
results are relatively insensitive to the exact list of theoretical
mass models.

%These excluded models include Brockmann et al.~\cite{DDMED}, Fuchs et
%al.~\cite{DDPC1}, Lalazissis et al.~\cite{NL3}~\cite{DDME2}, and
%Stoitsov et al. (2003)~\cite{SDNP03}.

%empirical formulae~\cite{Mumpower16} fit to the
%AME1995~\cite{AME1995} and AME2003~\cite{AME2003}.
%AME2012~\cite{AME2012}and AME2016~\cite{AME2016} are refinements to
%the earlier Atomic Mass Evaluations, so are also included in the
%calculation. Another series of forumlae are in with the
%%Hartree-Fock-Bogoliubov (HFB) approach~\cite{Mumpower16}. These
%formulae start from least to most complicated; the ones included in
%this calculation are the following select models:
%Moller-Nix-Myers-Swiatecki-Kratz (MNMSK)~\cite{Moller95} which is
%based on the Finite-Range Droplet Macroscopic model (FRDM95), and a
%microscopic-macroscopic nuclear mass model.

\begin{figure}
\includegraphics[width=0.23\textwidth]{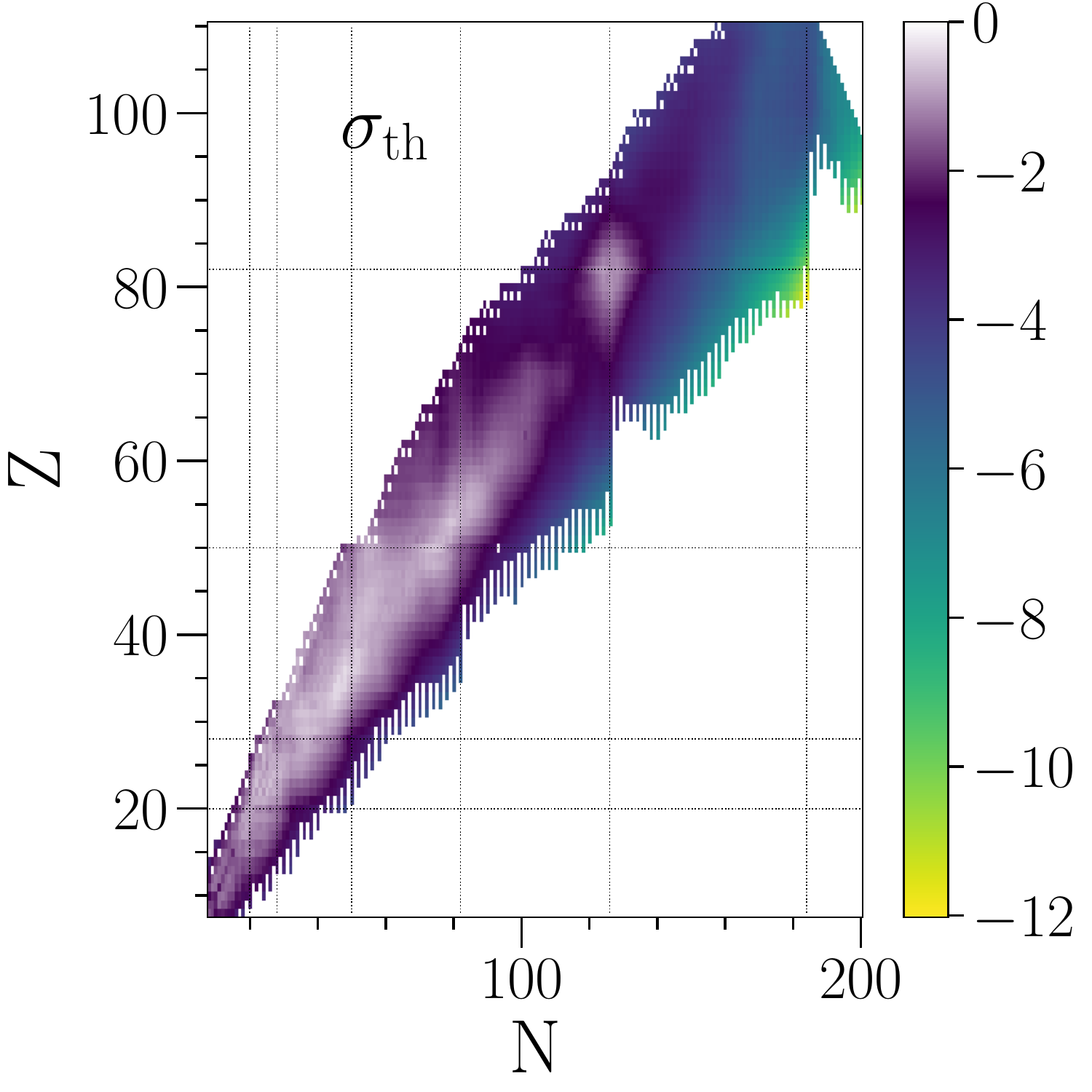}
\includegraphics[width=0.23\textwidth]{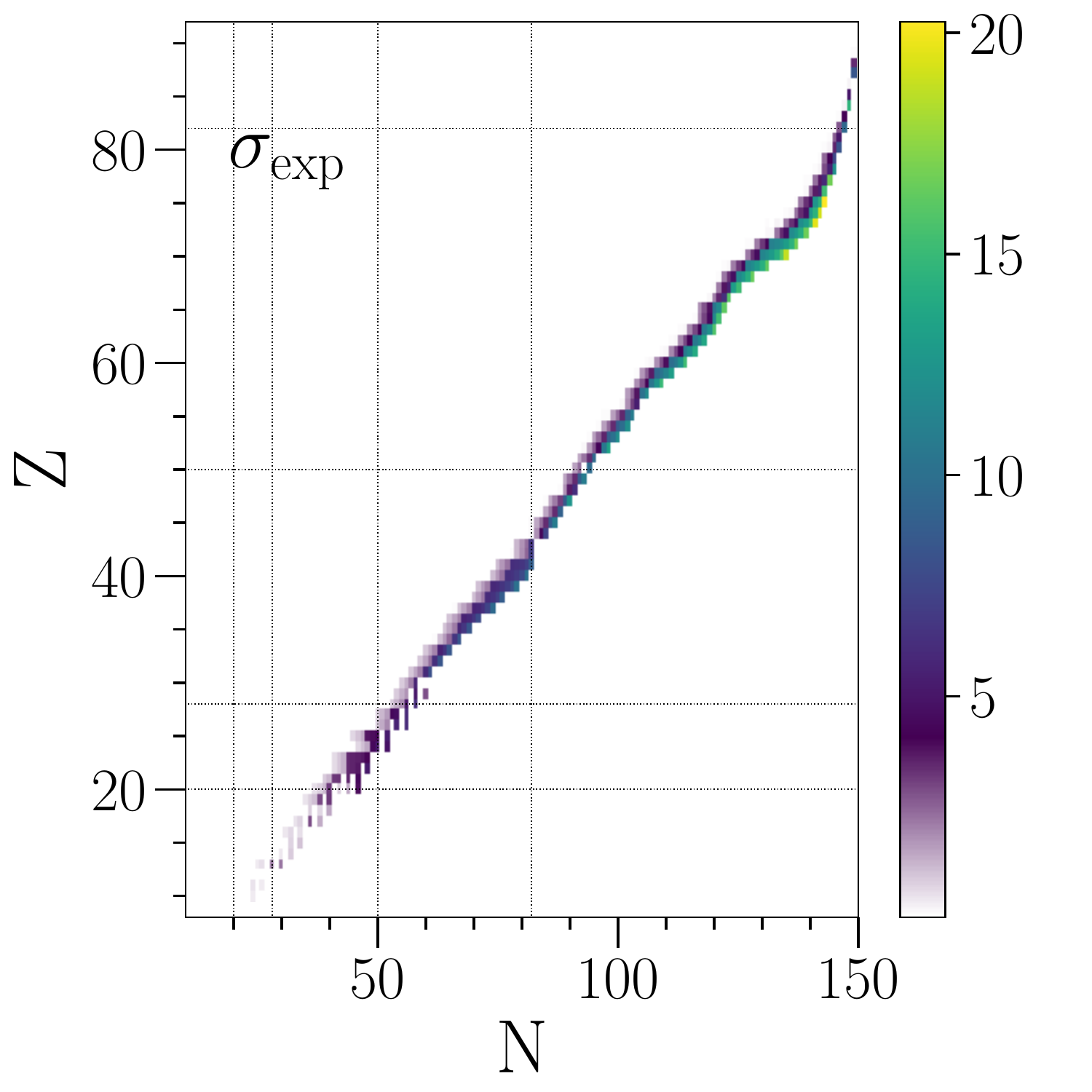}
\includegraphics[width=0.23\textwidth]{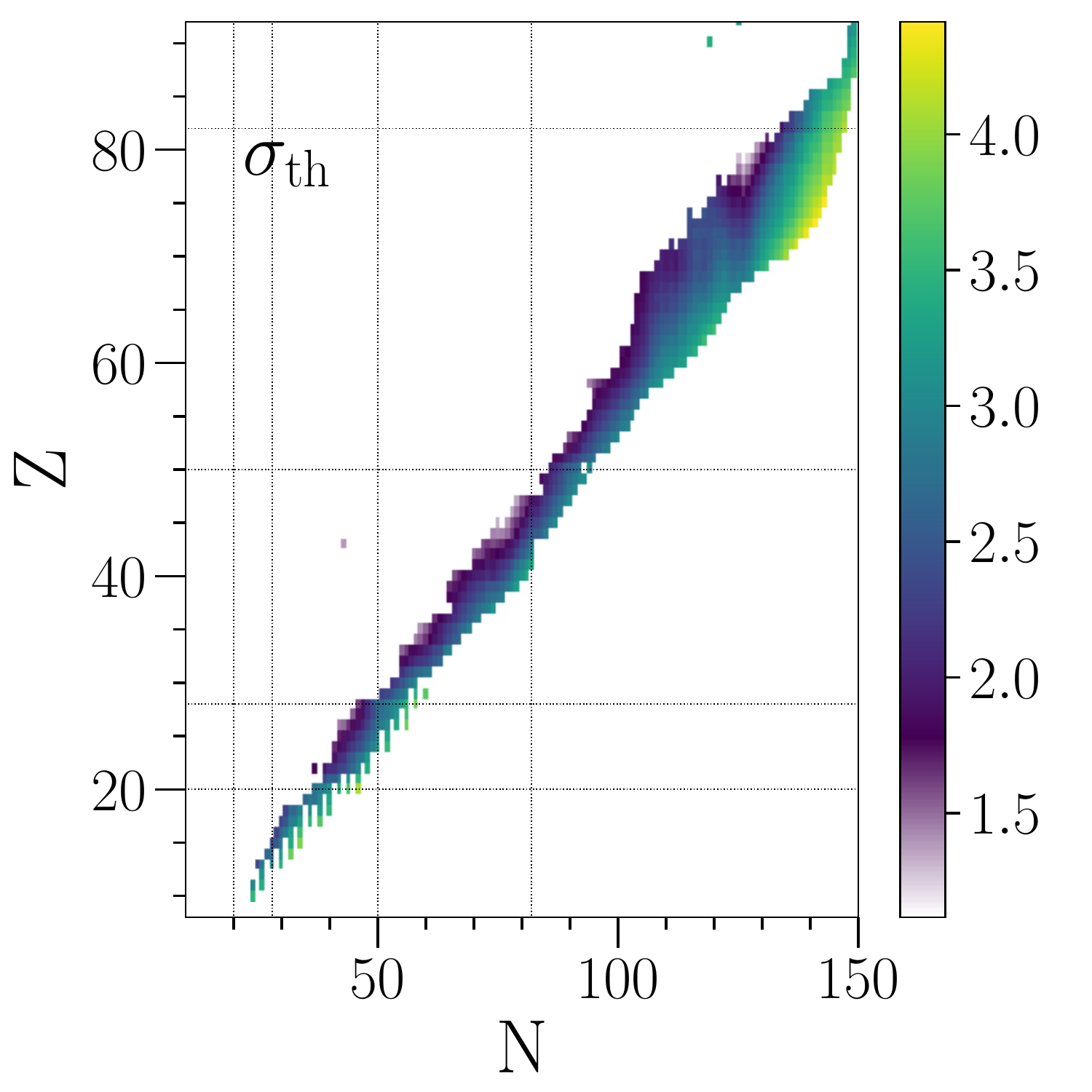}
\includegraphics[width=0.23\textwidth]{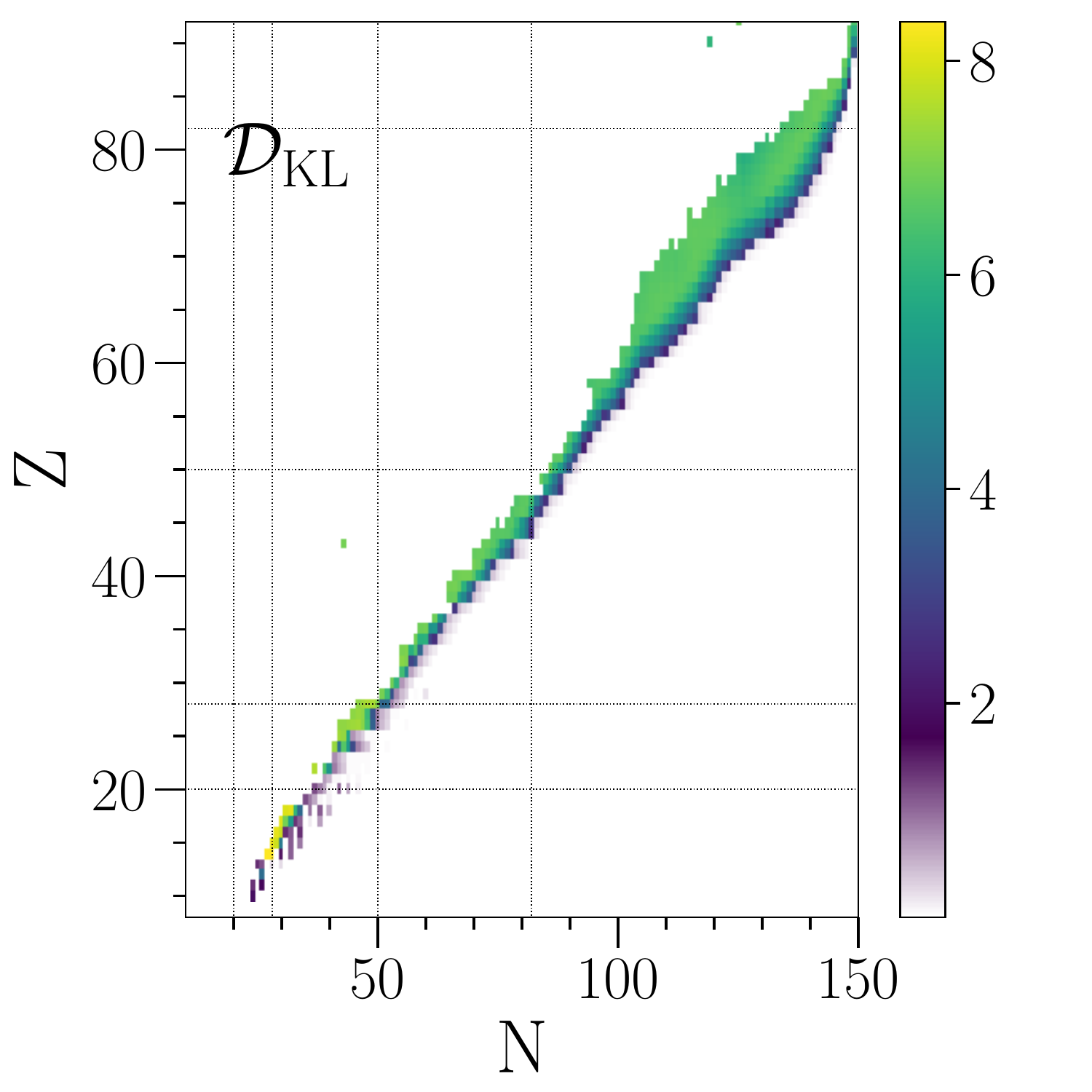}
\caption{The top-left panel shows the standard deviation of the
  theoretical mass model predictions over all nuclei contained in all
  of the models used in this work. The upper-right panel shows the
  anticipated uncertainty obtained in an FRIB mass measurement. The
  lower-left panel also shows the standard deviation of the
  theoretical mass model predictions, but now constrained only for
  those nuclei which are not yet, but could be, measured in FRIB. The
  lower-right panel shows the KL divergence for a mass measurement of
  each nucleus. The information obtained is larger as one approaches
  the valley of stability, except for light nuclei where a more
  complicated variation in A is shown. \label{fig:masses}}
\end{figure}

Using these mass models, the standard deviation in the predicted
nuclear mass, $\sigma_{\mathrm{th}}$ is plotted in the upper-left
panel of Fig~\ref{fig:masses}. (We include experimentally measured
nuclei in this plot, even though for these nuclei the variation across
mass models is best measured by the variation across experimental mass
measurements rather than from theory.) The largest uncertainties
between mass models occur at the edge of the neutron drip line very
close to the $Z=82$ shell. This result is not surprising; it is
well-known that theoretical mass models have a difficult time
correctly describing shell effects.
%This means that the farther away from stability, the mass models will
%have a larger variance from one another, and there will be a
%noticeable uncertainty in among the models.

Given a nucleus, we assume that the experimental information obtained
by a FRIB measurement is a Gaussian with a standard deviation of
\begin{equation}
    \sigma_{\mathrm{ex}}=\delta m \, .
    \label{eq:iex}
\end{equation}
This aligns with our intuitive expectation; more precise measurements
imply a smaller value for $\sigma_{\mathrm{ex}}$ and a more
strongly-peaked (i.e. more informative) probability distribution. The
experimental information, for those nuclei which are not already
experimentally measured in 2020 AME~\cite{Huang21ta}, is given in the
upper-right panel of Fig~\ref{fig:masses}. Clearly the information
obtained from the experiment is much larger closer to the valley of
stability because we can measure those nuclei with a greater
precision. The lower left panel of Fig.~\ref{fig:masses} shows the
value of $\sigma_{\mathrm{th}}$ for nuclei which are accessible from
FRIB. Many nuclei far from stability, could in principle strongly
constrain theoretical mass models but they cannot be easily created at
FRIB.
%Note that the trend of $\sigma_{\mathrm{th}}$ is reversed in the
%lower left panel from the upper left panel. This originates in the
%fact that FRIB can more easily extend to the neutron drip line in
%lighter nuclei than in heavier nuclei.

Presuming that the new mass measurement will result in the predicted
value, the information gain for a mass measurement given this estimate
of the utility is
$D_{\mathrm{KL}}(\sigma_{\mathrm{post}}|\sigma_{\mathrm{th}})$ (see
Eq.~\ref{eq:kldiv}), where $\sigma_{\mathrm{post}}$ represents the
uncertainty in the posterior probability distribution. The posterior
uncertainty, presuming a product of two Gaussian distributions, is
\begin{equation}
  \sigma_{\mathrm{post}}=\left(\frac{1}
      {\sigma_{\mathrm{th}^2}}+\frac{1}{\sigma_{\mathrm{exp}^2}}\right)^{-1/2} \, ,
    \label{eq:twog}
\end{equation}
and increases farther away from stability. The larger the value of $
\sigma_{\mathrm{ex}}$, the less information is gained from measuring
the mass of a particular nucleus. Eq.~\ref{eq:twog} ensures that an
experimental measurement never decreases our knowledge of a nucleus;
$\sigma_{\mathrm{post}}$ is always smaller than
$\sigma_{\mathrm{th}}$.

The value of $D_{\mathrm{KL}}$ is plotted in the lower right panel of
Fig~\ref{fig:masses}. The nuclei with the largest $D_{\mathrm{KL}}$
represent the mass measurements which provide the most information
relative to theoretical mass models which attempt to fit the entire
mass chart. Note that, for the tabulated nuclei with the largest
values of $D_{\mathrm{KL}}$, we found $\sigma_{\mathrm{ex}} \gg
\sigma_{\mathrm{th}}$ so $\sigma_{\mathrm{ex}} \approx
\sigma_{\mathrm{post}}$, but this is not true in for all of the
isotopes in out data set.

These results are also summarized in Table~\ref{tab:bdttab}, where the
ten nuclei with the largest values of $D_{\mathrm{KL}}$ are, in order,
$^{42}\mathrm{Si}$, $^{41}\mathrm{Si}$, $^{43}\mathrm{P}$,
$^{67}\mathrm{Mn}$, $^{59}\mathrm{Ti}$, $^{44}\mathrm{P}$,
$^{77}\mathrm{Ni}$, $^{45}\mathrm{S}$, $^{68}\mathrm{Mn}$, and
$^{47}\mathrm{Cl}$. The properties of the FRIB instrumentation
dominate the information gain, because Table~\ref{tab:bdttab} shows
that the largest value of $D_{\mathrm{KL}}$ is strongly correlated
with the smallest value of $\sigma_{\mathrm{ex}}$. The more accurate
the experiment, the more information we gain. However, there are
exceptions to this rule. The nucleus with the minimum value of
$\sigma_{\mathrm{ex}}$, $^{41}\mathrm{Si}$, is not the nucleus with
the maximum value of $D_{\mathrm{KL}}$, $^{42}\mathrm{Si}$. This
occurs when $\sigma_{\mathrm{th}}$ is sufficiently large, for nearly
equal values of $\sigma_{\mathrm{ex}}$. In this case, the theoretical
uncertainty for $^{42}\mathrm{Si}$ is larger because theoretical
models have a difficult time accurately describing the $N=28$ closed
shell. When $\sigma_{\mathrm{ex}}$ is nearly equal between two nuclei,
we gain the most information by measuring the nucleus with the larger
theoretical uncertainty, i.e. a larger value of
$\sigma_{\mathrm{th}}$. Both the properties of the FRIB facility and
the theoretical mass models impact the information gain, but because
the range of values of $\sigma_{\mathrm{ex}}$ is much larger than the
range of values of $\sigma_{\mathrm{th}}$, the properties of the FRIB
facility have the strongest impact. This is not necessarily the case
with the r-process results we show below.
 
\begin{table}
%\setlength{\arrayrulewidth}{0.4mm}
%\setlength{\tabcolsep}{18pt}
%\renewcommand{\arraystretch}{1.5}
%\centering
\begin{tabular}{p{1cm}p{1cm}llp{1cm}lp{1cm}l}
%\hline
%\multicolumn{4}{|c|}{Maximum Information Gain Relative to Theoretical Mass Models} \\
%\hline
 Isotope & Z & N & $\sigma_{\mathrm{ex}}$ & $\sigma_{\mathrm{th}}$ & $\sigma_{\mathrm{post}}$ & $D_{\mathrm{KL}}$\\
 \hline
$^{42}$Si & 14 & 28 & 3.94$\times10^{-4}$ & 2.78 & 3.94$\times10^{-4}$ & 8.36 \\ 
$^{41}$Si & 14 & 27 & 3.82$\times10^{-4}$ & 2.63 & 3.82$\times10^{-4}$ & 8.34 \\ 
$^{43}$P  & 15 & 28 & 4.01$\times10^{-4}$ & 2.39 & 4.01$\times10^{-4}$ & 8.19 \\ 
$^{45}$S  & 16 & 29 & 4.19$\times10^{-4}$ & 2.30 & 4.19$\times10^{-4}$ & 8.11 \\ 
$^{50}$Ar & 18 & 32 & 4.92$\times10^{-4}$ & 2.66 & 4.92$\times10^{-4}$ & 8.10 \\ 
$^{47}$Cl & 17 & 30 & 4.41$\times10^{-4}$ & 2.33 & 4.41$\times10^{-4}$ & 8.07 \\ 
$^{49}$Ar & 18 & 31 & 4.58$\times10^{-4}$ & 2.29 & 4.58$\times10^{-4}$ & 8.02 \\ 
$^{44}$P  & 15 & 29 & 5.94$\times10^{-4}$ & 2.51 & 5.94$\times10^{-4}$ & 7.85 \\ 
$^{46}$S  & 16 & 30 & 6.42$\times10^{-4}$ & 2.63 & 6.42$\times10^{-4}$ & 7.82 \\ 
$^{59}$Ti & 22 & 37 & 5.62$\times10^{-4}$ & 1.77 & 5.62$\times10^{-4}$ & 7.56 \\ 
$^{76}$Ni & 28 & 48 & 7.08$\times10^{-4}$ & 2.11 & 7.08$\times10^{-4}$ & 7.50 \\ 
$^{77}$Ni & 28 & 49 & 8.19$\times10^{-4}$ & 2.41 & 8.19$\times10^{-4}$ & 7.49 \\ 
$^{72}$Fe & 26 & 46 & 6.93$\times10^{-4}$ & 1.92 & 6.93$\times10^{-4}$ & 7.43 \\ 
$^{67}$Mn & 25 & 42 & 6.26$\times10^{-4}$ & 1.69 & 6.26$\times10^{-4}$ & 7.40 \\ 
$^{75}$Ni & 28 & 47 & 6.98$\times10^{-4}$ & 1.88 & 6.98$\times10^{-4}$ & 7.40 \\ 
$^{71}$Fe & 26 & 45 & 6.64$\times10^{-4}$ & 1.75 & 6.64$\times10^{-4}$ & 7.38 \\ 
$^{65}$Cr & 24 & 41 & 6.59$\times10^{-4}$ & 1.67 & 6.59$\times10^{-4}$ & 7.34 \\ 
$^{73}$Co & 27 & 46 & 6.89$\times10^{-4}$ & 1.74 & 6.89$\times10^{-4}$ & 7.34 \\ 
$^{70}$Fe & 26 & 44 & 6.53$\times10^{-4}$ & 1.65 & 6.53$\times10^{-4}$ & 7.33 \\ 
$^{68}$Mn & 25 & 43 & 6.97$\times10^{-4}$ & 1.73 & 6.97$\times10^{-4}$ & 7.31 \\

%\hline
\end{tabular}
\caption{The maximum information gain for experimental
nuclear mass measurements relative to the information in theoretical mass models, as measured by the KL divergence. \label{tab:bdttab}}
\end{table}

%All of the aforementioned isotopes are at a distance in between the
%dripline and valley of stability, but are closer to the dripline.
%FRIB is proposed to produce these exotic isotopes at higher yields
%than at previous facilities. These heavy isotopes can now be studied
%in a way that they have not been before, simply due to the amount
%being produced.

%Other properties of nuclei of interest include further study of
%nuclear shell effects. For instance, effects around $Z=14$ exhibit
%strange behavior, and studying these nuclei can give rise to further
%understanding nuclear structure and the shell model.

\section{Information Gain Relative to r-Process Abundances}

%As stated earlier, there is much uncertainty around thoroughly
%understanding the creation of these exotic elements due to them being
%far from stability. Understanding nuclear masses and beta decay
%properties, which will be analyzed in FRIB, are being improved every
%year. A relatively new model is the finite-range droplet model,
%FRDM(2012), which is more accurate than the previous AME2003
%\cite{Kratz}.

The rapid neutron capture process, or r-process, is responsible for
the heaviest elements in the universe. Under varying astrophysical
conditions, nuclei can rapidly capture neutrons, creating heavy
isotopes not possible via fusion in stars. The origin of r-process
nuclei is not yet understood; experimental mass measurements are
critical in determining the astrophysical site of the
r-process~\cite{Cowan}. Core-collapse supernovae and neutron star
mergers may both contribute to the observed r-process
abundances~\cite{Thielemann}. These r-process sites listed have
varying physical conditions which lead to different abundance
patterns.
%All of these possible r-process sites listed have distinct conditions which should lead to specific abundances: initial chemical composition,initial neutron density and richness, and temperature and density throughout time 
In core-collapse supernovae, r-process nucleosynthesis occurs in the
neutrino-driven wind above the newly-born ``protoneutron'' star. If
this neutrino-driven wind does not move too quickly (i.e. if the
dynamical timescale is not too short), then the nuclear reactions
maintain a robust $(n,\gamma)\rightleftarrows(\gamma,n)$ equilibrium,
referred to as a ``hot wind''. During this hot wind, also occurring at
high temperatures as the name suggests, the rapid neutron capture
process happens quicker than $\mathrm{\beta^{-}}$ decay. If the wind
moves faster and this equilibrium is not achieved, and is also at
relatively low temperatures, then this scenario is referred to as a
``cold wind'', where the neutron capture and $\mathrm{\beta^{-}}$
decays happen at a comparable rate \cite{Arcones2011}. In the hot wind
scenario, r-process nucleosynthesis may proceed with either a high
($S>120$) or low ($S<120$) entropy, which would yield differing peaks
in isotope production~\cite{Mumpower16}. Another method for heavy
element production is neutron star mergers, but unlike with the winds
of a core-collapse supernova, they produce nuclei through fission
recycling~\cite{Yong}. This process is dominated by neutron capture
and beta decay, and spends almost no time in the
$(n,\gamma)\rightleftarrows(\gamma,n)$ equilibrium phase. The
composition is so neutron rich, more so than the other three
processes, that it quickly reaches the neutron drip
line~\cite{Cote2019}. We will refer to the high and low entropy hot
winds as HEHW and LEHW, respectively, the cold wind as CW, and neutron
star mergers as NSM.
 
%neutrino-driven wind which can lead to nuclei as heavy as thorium~\cite{X}. 

In order to study these previously described processes, we must find
which experiments will contribute the most knowledge to our current
understanding of astrophysical systems. To calculate the theoretical
information, we use
\begin{equation}
    \sigma_{\mathrm{th}}=F \sigma_0 \, ,
\end{equation}
where $F$ is the integrated absolute difference in isobaric mass
fractions between a baseline calculation and calculation with modified
input nuclear data, as tabulated in Ref.~\cite{Mumpower16} and
$\sigma_0 \equiv 1~\mathrm{MeV}$. An arbitrary constant is necessary
because there is no unique way of determining which probability
distribution one should use to determine the information. However, our
results are only weakly-dependent on this choice, because the KL
divergence depends only on the ratio
$\sigma_{\mathrm{post}}/\sigma_{\mathrm{th}}$.

The most informative nuclei are found by once again using the KL
divergence for the four astrophysical processes. These are found using
the same KL divergence from Eq.~\ref{eq:kldiv} for the LEHW, HEHW, CW,
and NSM, with the notation taken in this paper to be as follows for
each of the astrophysical processes' KL divergences:
$D_{\mathrm{LEHW}}$, $D_{\mathrm{HEHW}}$, $D_{\mathrm{CW}}$, and
$D_{\mathrm{NSM}}$.

\begin{table}
%\setlength{\arrayrulewidth}{0.4mm}
%\setlength{\tabcolsep}{18pt}
%\renewcommand{\arraystretch}{1.5}
%\centering
\begin{tabular}{p{1cm}p{1cm}lp{1cm}lp{1cm}l}
%\hline
%\multicolumn{4}{|c|}{Maximum Information Gain Relative to r-Process Abundances (LEHW)} \\
%\hline
 Isotope & Z & N & $\sigma_{\mathrm{th}}$ & $\sigma_{\mathrm{post}}$ & $D_{\mathrm{LEHW}}$\\
 \hline
$^{133}$In & 49 & 84  & 9.44 & 1.27$\times10^{-3}$ & 8.41  \\ 
$^{134}$In & 49 & 85  & 9.35 & 2.16$\times10^{-3}$ & 7.87  \\ 
$^{134}$Cd & 48 & 86  & 52.4 & 0.0126             & 7.83  \\ 
$^{133}$Cd & 48 & 85  & 23.4 & 5.72$\times10^{-3}$ & 7.82  \\ 
$^{135}$In & 49 & 86  & 10.5 & 4.66$\times10^{-3}$ & 7.22  \\ 
$^{196}$Hf & 72 & 124 & 8.93 & 4.50$\times10^{-3}$ & 7.09  \\ 
$^{136}$Sn & 50 & 86  & 2.37 & 1.32$\times10^{-3}$ & 7.00  \\ 
$^{128}$Ag & 47 & 81  & 2.22 & 1.24$\times10^{-3}$ & 6.99  \\ 
$^{141}$Te & 52 & 89  & 2.18 & 1.37$\times10^{-3}$ & 6.88  \\ 
$^{128}$Pd & 46 & 82  & 4.71 & 4.23$\times10^{-3}$ & 6.52  \\ 
$^{127}$Pd & 46 & 81  & 1.49 & 1.44$\times10^{-3}$ & 6.44  \\ 
$^{130}$Ag & 47 & 83  & 12.5 & 0.0129             & 6.38  \\ 
$^{129}$Ag & 47 & 82  & 1.92 & 2.16$\times10^{-3}$ & 6.29  \\ 
$^{166}$Sm & 62 & 104 & 1.50 & 1.82$\times10^{-3}$ & 6.22  \\ 
$^{137}$Sn & 50 & 87  & 2.17 & 2.74$\times10^{-3}$ & 6.18  \\ 
$^{138}$Sn & 50 & 88  & 4.55 & 5.82$\times10^{-3}$ & 6.16  \\ 
$^{197}$Hf & 72 & 125 & 4.85 & 6.29$\times10^{-3}$ & 6.15  \\ 
$^{156}$Ce & 58 & 98  & 1.29 & 1.69$\times10^{-3}$ & 6.14  \\ 
$^{194}$Hf & 72 & 122 & 1.96 & 2.60$\times10^{-3}$ & 6.12  \\ 
$^{162}$Nd & 60 & 102 & 1.71 & 2.33$\times10^{-3}$ & 6.10  \\ 

%\hline
\end{tabular}
\caption{The maximum information gain for experimental nuclear mass
  measurements relative to the information in the low entropy hot wind
  r-process astrophysical models, as measured by the KL
  divergence. \label{tab:lehwtab}}
\end{table}

\begin{table}
%\setlength{\arrayrulewidth}{0.4mm}
%\setlength{\tabcolsep}{18pt}
%\renewcommand{\arraystretch}{1.5}
%\centering
\begin{tabular}{p{1cm}p{1cm}lp{1cm}lp{1cm}l}
%\hline
%\multicolumn{4}{|c|}{Maximum Information Gain Relative to r-Process Abundances (HEHW)} \\
%\hline
 Isotope & Z & N & $\sigma_{\mathrm{th}}$ & $\sigma_{\mathrm{post}}$ & $D_{\mathrm{HEHW}}$\\
 \hline
$^{136}$Sn & 50 & 86  & 83.7 & 1.32$\times10^{-3}$ & 10.6  \\
$^{133}$In & 49 & 84  & 74.6 & 1.27$\times10^{-3}$ & 10.5  \\
$^{134}$In & 49 & 85  & 72.8 & 2.16$\times10^{-3}$ & 9.93  \\
$^{135}$In & 49 & 86  & 73.8 & 4.66$\times10^{-3}$ & 9.17  \\
$^{138}$Sb & 51 & 87  & 19.2 & 1.30$\times10^{-3}$ & 9.10  \\ 
$^{139}$Sb & 51 & 88  & 17.3 & 1.41$\times10^{-3}$ & 8.92  \\ 
$^{138}$Sn & 50 & 88  & 29.6 & 5.82$\times10^{-3}$ & 8.04  \\ 
$^{136}$In & 49 & 87  & 71.0 & 0.0154              & 7.93  \\ 
$^{196}$Hf & 72 & 124 & 13.7 & 4.50$\times10^{-3}$ & 7.52  \\ 
$^{141}$Te & 52 & 89  & 4.13 & 1.37$\times10^{-3}$ & 7.51  \\ 
$^{140}$Sb & 51 & 89  & 14.1 & 4.70$\times10^{-3}$ & 7.51  \\ 
$^{128}$Ag & 47 & 81  & 3.51 & 1.24$\times10^{-3}$ & 7.45  \\ 
$^{165}$Sm & 62 & 103 & 4.04 & 1.61$\times10^{-3}$ & 7.32  \\ 
$^{167}$Eu & 63 & 104 & 3.95 & 1.59$\times10^{-3}$ & 7.32  \\ 
$^{194}$Hf & 72 & 122 & 5.74 & 2.60$\times10^{-3}$ & 7.20  \\ 
$^{197}$Ta & 73 & 124 & 5.31 & 2.46$\times10^{-3}$ & 7.18  \\ 
$^{166}$Sm & 62 & 104 & 3.77 & 1.82$\times10^{-3}$ & 7.14  \\ 
$^{197}$Hf & 72 & 125 & 12.1 & 6.29$\times10^{-3}$ & 7.06  \\ 
$^{170}$Gd & 64 & 106 & 2.98 & 1.63$\times10^{-3}$ & 7.01  \\ 
$^{168}$Eu & 63 & 105 & 3.17 & 1.75$\times10^{-3}$ & 7.00  \\

%\hline
\end{tabular}
\caption{The maximum information gain for experimental nuclear mass
  measurements relative to the information in the high entropy hot
  wind r-process astrophysical models, as measured by the KL
  divergence. \label{tab:hehwtab}}
\end{table}

\begin{table}
%\setlength{\arrayrulewidth}{0.4mm}
%\setlength{\tabcolsep}{18pt}
%\renewcommand{\arraystretch}{1.5}
%\centering
\begin{tabular}{p{1cm}p{1cm}lp{1cm}lp{1cm}l}
%\hline
%\multicolumn{4}{|c|}{Maximum Information Gain Relative to r-Process Abundances (CW)} \\
%\hline
 Isotope & Z & N & $\sigma_{\mathrm{th}}$ & $\sigma_{\mathrm{post}}$ & $D_{\mathrm{CW}}$\\
 \hline
$^{133}$In & 49 & 84  & 4.20 & 1.27$\times10^{-3}$ & 7.60  \\ 
$^{141}$Te & 52 & 89  & 3.44 & 1.37$\times10^{-3}$ & 7.33  \\ 
$^{134}$In & 49 & 85  & 4.12 & 2.16$\times10^{-3}$ & 7.05  \\ 
$^{143}$I  & 53 & 90  & 1.38 & 1.34$\times10^{-3}$ & 6.43  \\ 
$^{128}$Ag & 47 & 81  & 1.23 & 1.24$\times10^{-3}$ & 6.40  \\ 
$^{142}$Te & 52 & 90  & 2.22 & 2.37$\times10^{-3}$ & 6.34  \\ 
$^{138}$Sb & 51 & 87  & 0.89 & 1.30$\times10^{-3}$ & 6.03  \\ 
$^{136}$Sn & 50 & 86  & 0.89 & 1.32$\times10^{-3}$ & 6.02  \\ 
$^{139}$Sb & 51 & 88  & 0.85 & 1.41$\times10^{-3}$ & 5.90  \\ 
$^{130}$Pd & 46 & 84  & 43.7 & 0.0776              & 5.83  \\ 
$^{129}$Ag & 47 & 82  & 1.18 & 2.16$\times10^{-3}$ & 5.80  \\ 
$^{135}$In & 49 & 86  & 2.32 & 4.66$\times10^{-3}$ & 5.71  \\ 
$^{126}$Pd & 46 & 80  & 0.60 & 1.21$\times10^{-3}$ & 5.71  \\ 
$^{128}$Pd & 46 & 82  & 1.91 & 4.23$\times10^{-3}$ & 5.61  \\ 
$^{167}$Eu & 63 & 104 & 0.67 & 1.59$\times10^{-3}$ & 5.54  \\ 
$^{134}$Cd & 48 & 86  & 5.17 & 0.0126              & 5.52  \\ 
$^{194}$Hf & 72 & 122 & 1.03 & 2.60$\times10^{-3}$ & 5.48  \\
$^{192}$Hf & 72 & 120 & 0.71 & 1.82$\times10^{-3}$ & 5.47  \\ 
$^{154}$Ce & 58 & 96  & 0.55 & 1.46$\times10^{-3}$ & 5.43  \\ 
$^{193}$Hf & 72 & 121 & 0.71 & 1.97$\times10^{-3}$ & 5.39  \\ 
 
%\hline
\end{tabular}
\caption{The maximum information gain for experimental nuclear mass
  measurements relative to the information in the cold wind r-process
  astrophysical models, as measured by the KL
  divergence. \label{tab:cwtab}}
\end{table}

\begin{table}
%\setlength{\arrayrulewidth}{0.4mm}
%\setlength{\tabcolsep}{18pt}
%\renewcommand{\arraystretch}{1.5}
%\centering
\begin{tabular}{p{1cm}p{1cm}lp{1cm}lp{1cm}l}
%\hline
%\multicolumn{4}{|c|}{Maximum Information Gain Relative to r-Process Abundances (NSM)} \\
%\hline
 Isotope & Z & N & $\sigma_{\mathrm{th}}$ & $\sigma_{\mathrm{post}}$ & $D_{\mathrm{NSM}}$\\
 \hline
$^{133}$In & 49 & 84  & 6.07 & 1.27$\times10^{-3}$ & 7.97  \\ 
$^{134}$In & 49 & 85  & 5.92 & 2.16$\times10^{-3}$ & 7.42  \\ 
$^{128}$Ag & 47 & 81  & 2.89 & 1.24$\times10^{-3}$ & 7.26  \\ 
$^{126}$Pd & 46 & 80  & 2.46 & 1.21$\times10^{-3}$ & 7.12  \\ 
$^{127}$Ag & 47 & 80  & 1.68 & 1.18$\times10^{-3}$ & 6.76  \\ 
$^{129}$Ag & 47 & 82  & 2.90 & 2.16$\times10^{-3}$ & 6.70  \\ 
$^{127}$Pd & 46 & 81  & 1.85 & 1.44$\times10^{-3}$ & 6.66  \\ 
$^{124}$Pd & 46 & 78  & 1.34 & 1.16$\times10^{-3}$ & 6.56  \\ 
$^{126}$Ag & 47 & 79  & 1.28 & 1.17$\times10^{-3}$ & 6.49  \\ 
$^{141}$Te & 52 & 89  & 1.46 & 1.37$\times10^{-3}$ & 6.47  \\ 
$^{128}$Pd & 46 & 82  & 3.06 & 4.23$\times10^{-3}$ & 6.08  \\ 
$^{125}$Pd & 46 & 79  & 0.71 & 1.17$\times10^{-3}$ & 5.91  \\ 
$^{143}$I  & 53 & 90  & 0.63 & 1.34$\times10^{-3}$ & 5.65  \\ 
$^{142}$Te & 52 & 90  & 1.05 & 2.37$\times10^{-3}$ & 5.59  \\ 
$^{192}$Hf & 72 & 120 & 0.66 & 1.82$\times10^{-3}$ & 5.39  \\ 
$^{193}$Hf & 72 & 121 & 0.70 & 1.97$\times10^{-3}$ & 5.38  \\ 
$^{194}$Hf & 72 & 122 & 0.91 & 2.60$\times10^{-3}$ & 5.36  \\ 
$^{130}$Pd & 46 & 84  & 24.2 & 0.0776              & 5.24  \\ 
$^{135}$In & 49 & 86  & 1.45 & 4.66$\times10^{-3}$ & 5.24  \\ 
$^{154}$Ce & 58 & 96  & 0.45 & 1.46$\times10^{-3}$ & 5.23  \\ 
%\hline
\end{tabular}
\caption{The maximum information gain for experimental nuclear mass
  measurements relative to the information in the neutron star merger
  r-process astrophysical models, as measured by the KL
  divergence. \label{tab:nsmtab}}
\end{table}

\begin{figure}
\includegraphics[width=0.23\textwidth]{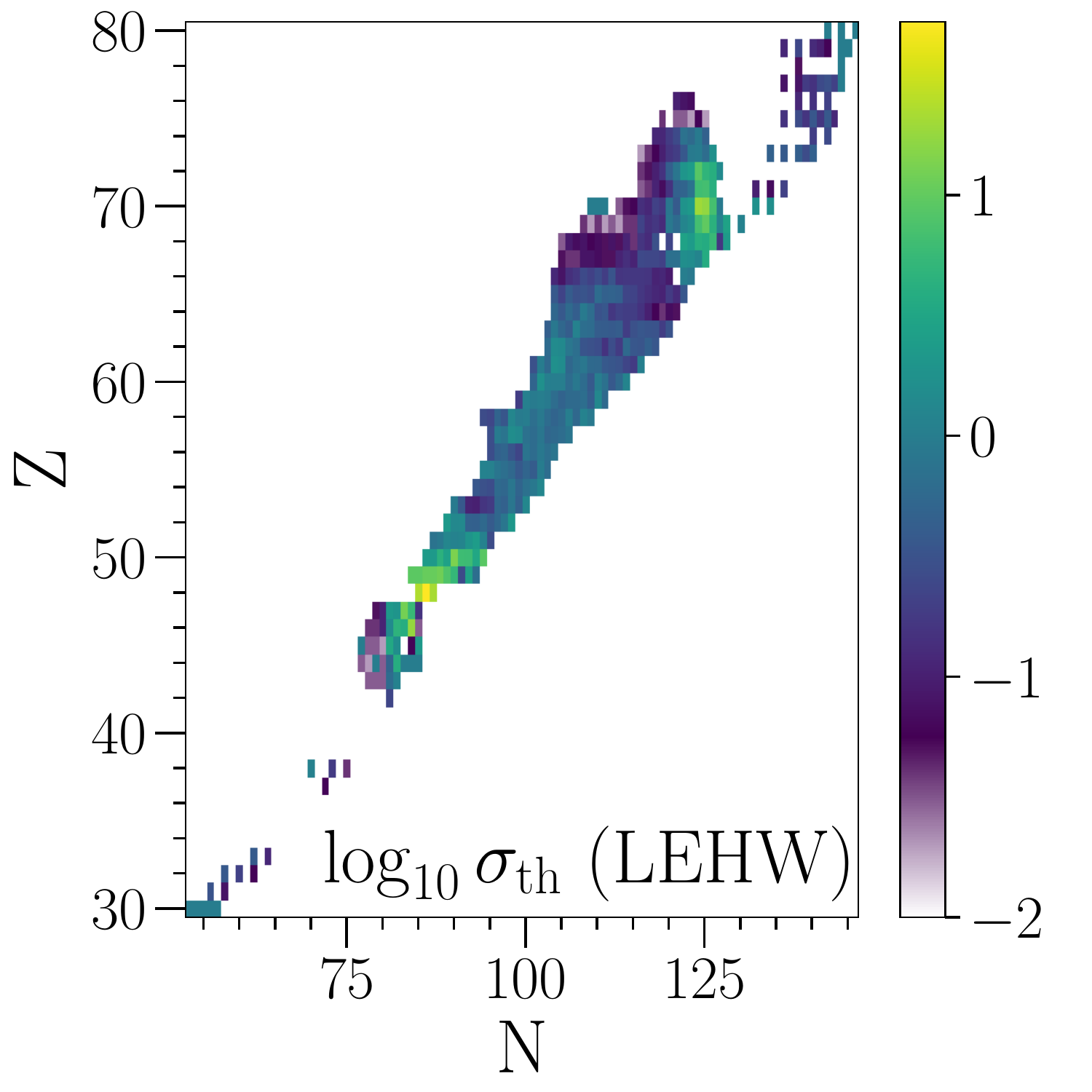}
\includegraphics[width=0.23\textwidth]{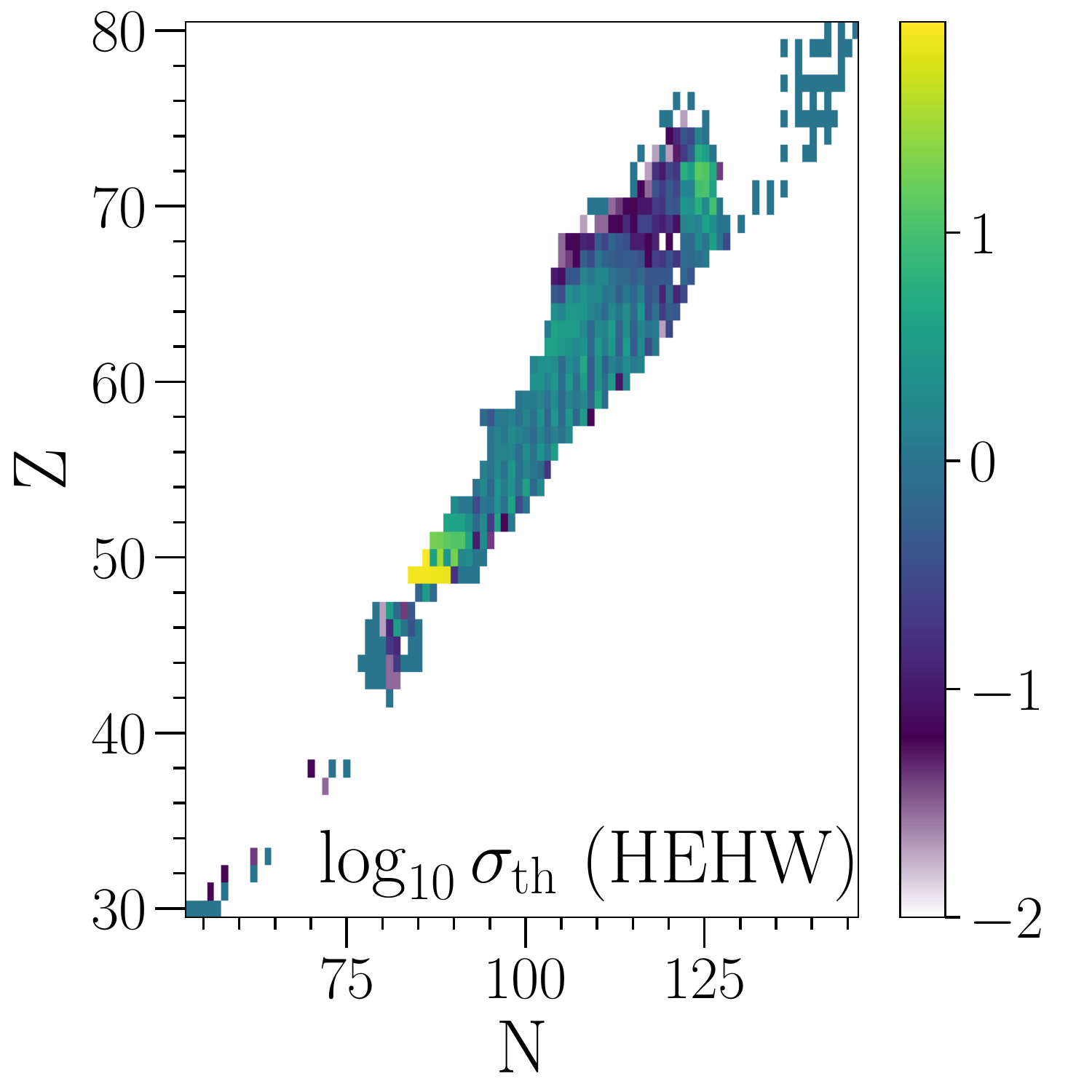}
\includegraphics[width=0.23\textwidth]{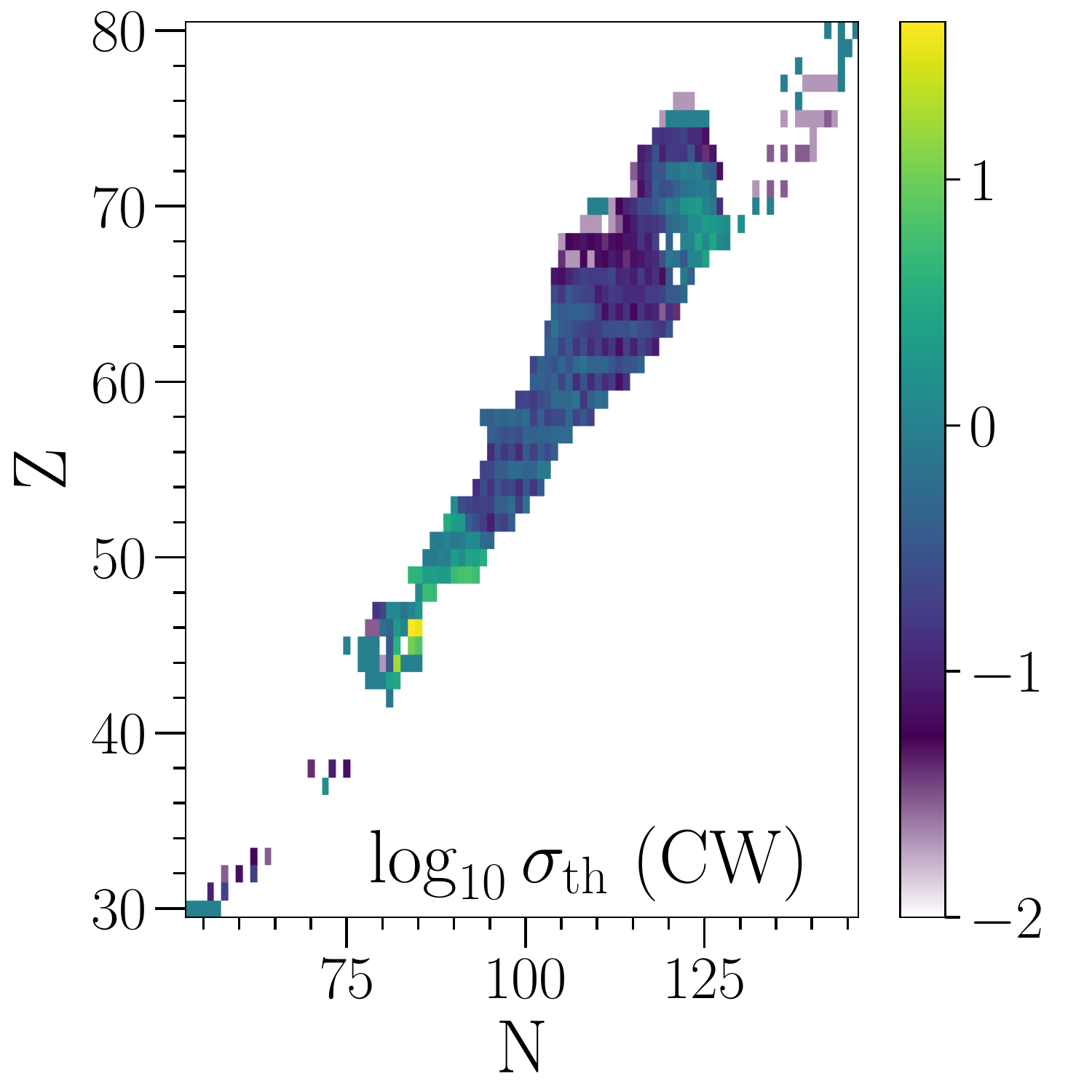}
\includegraphics[width=0.23\textwidth]{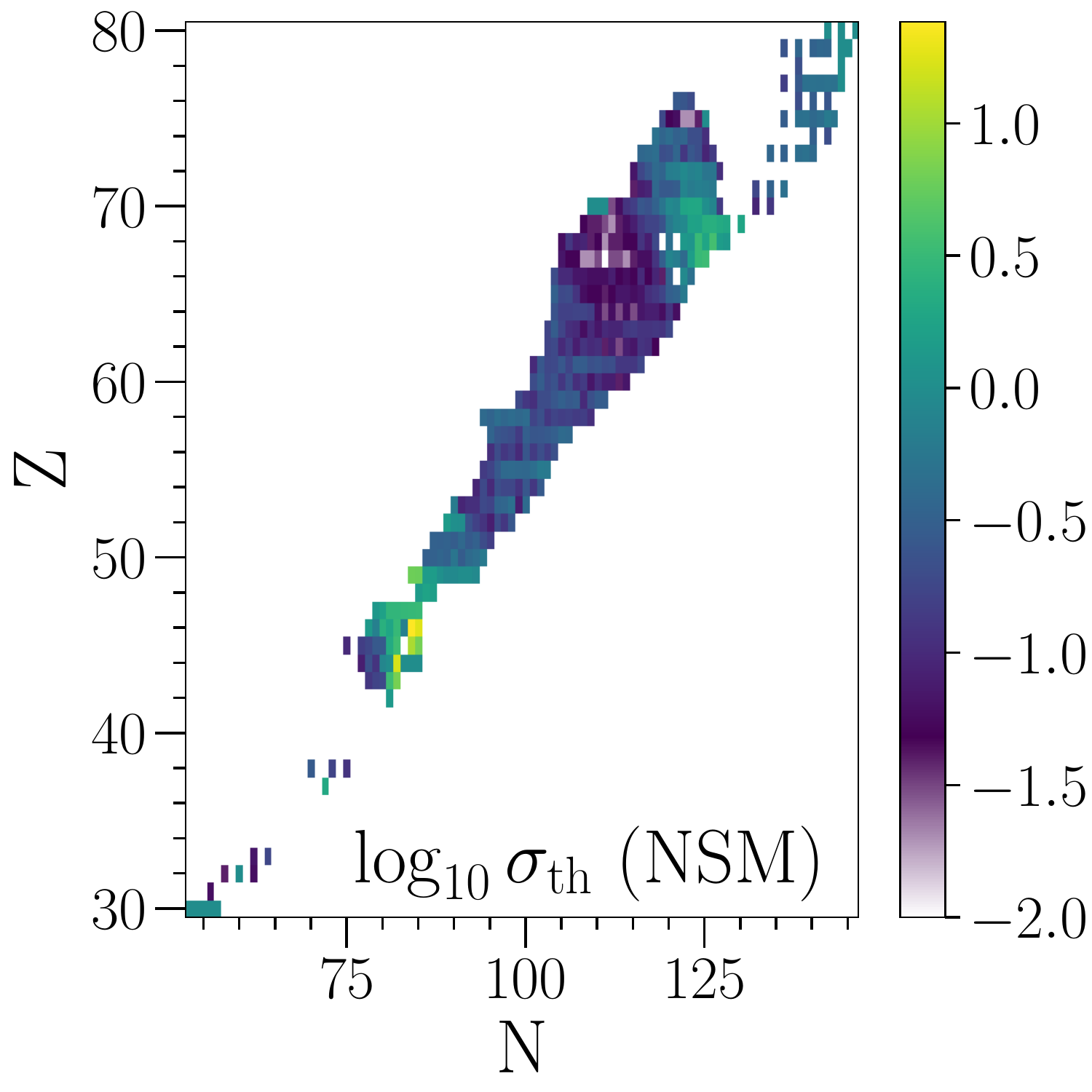}
\caption{Log of theoretical standard deviation for low entropy hot
  wind (top left). Log of theoretical standard deviation for high
  entropy hot wind (top right). Log of theoretical standard deviation
  for cold wind (bottom left). Log of theoretical standard deviation
  for neutron star mergers (bottom right). Nuclei in yellow represent
  the largest standard deviation, while nuclei with a smaller standard
  deviation are colored light purple.\label{fig:logth}}
\end{figure}

\begin{figure}
\includegraphics[width=0.23\textwidth]{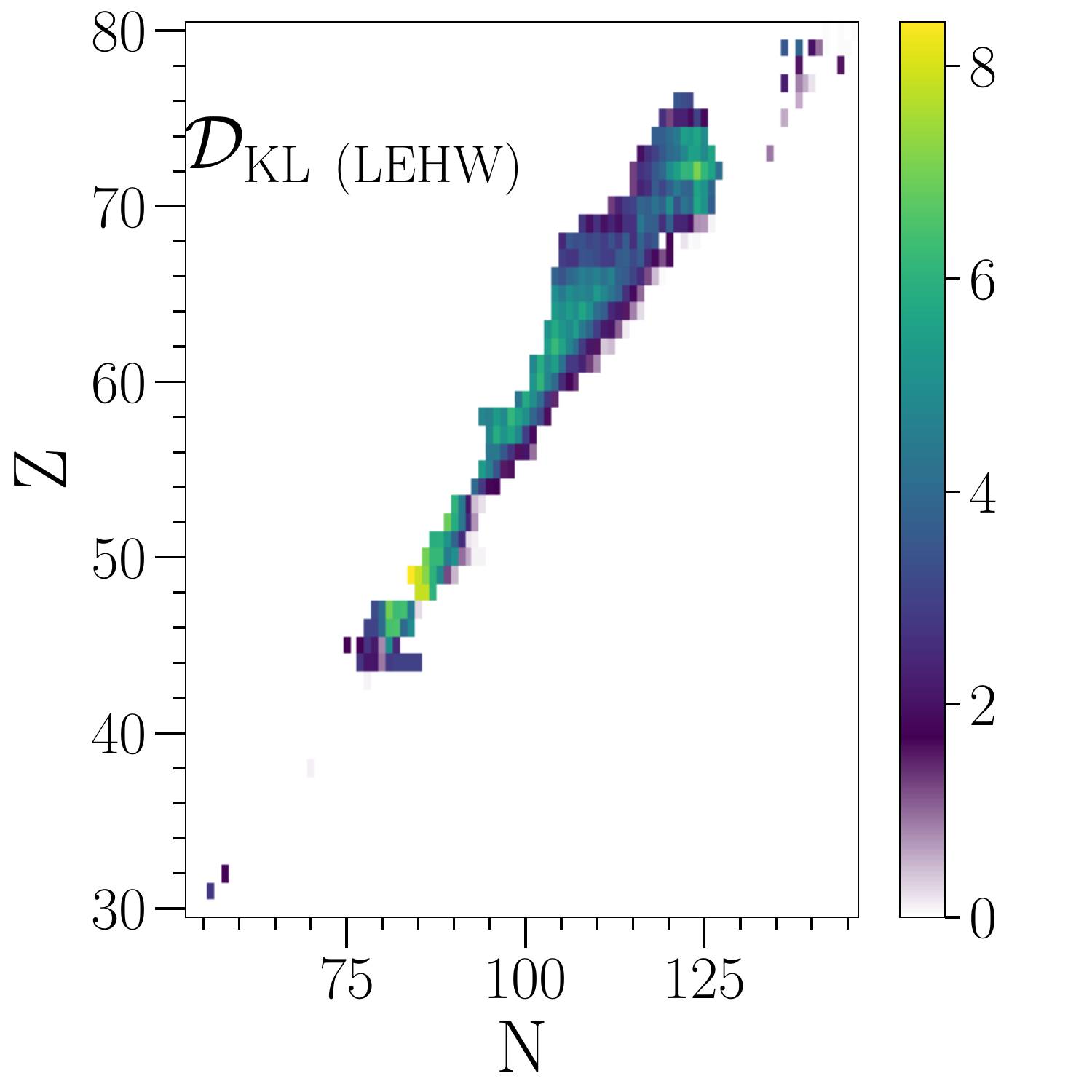}
\includegraphics[width=0.23\textwidth]{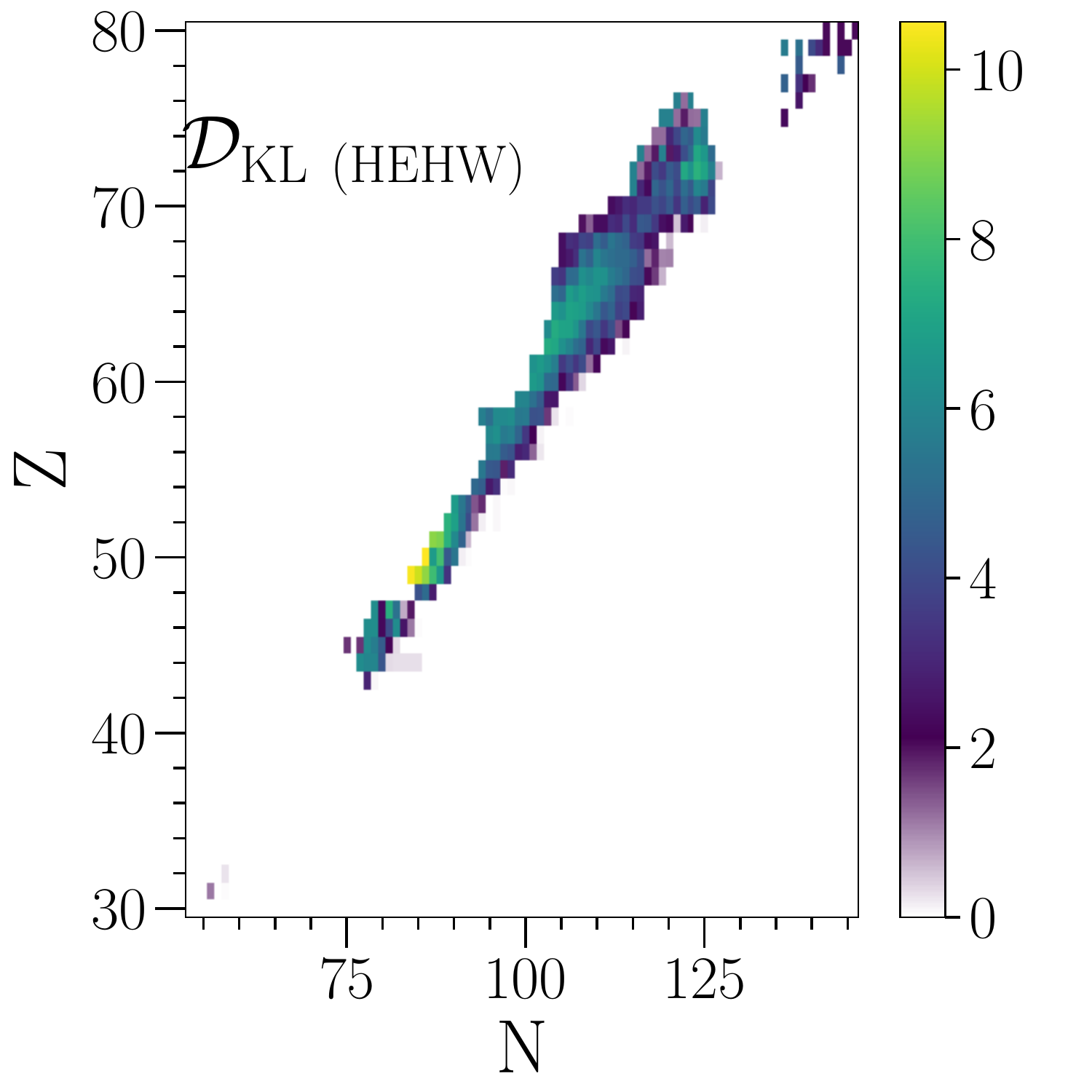}
\includegraphics[width=0.23\textwidth]{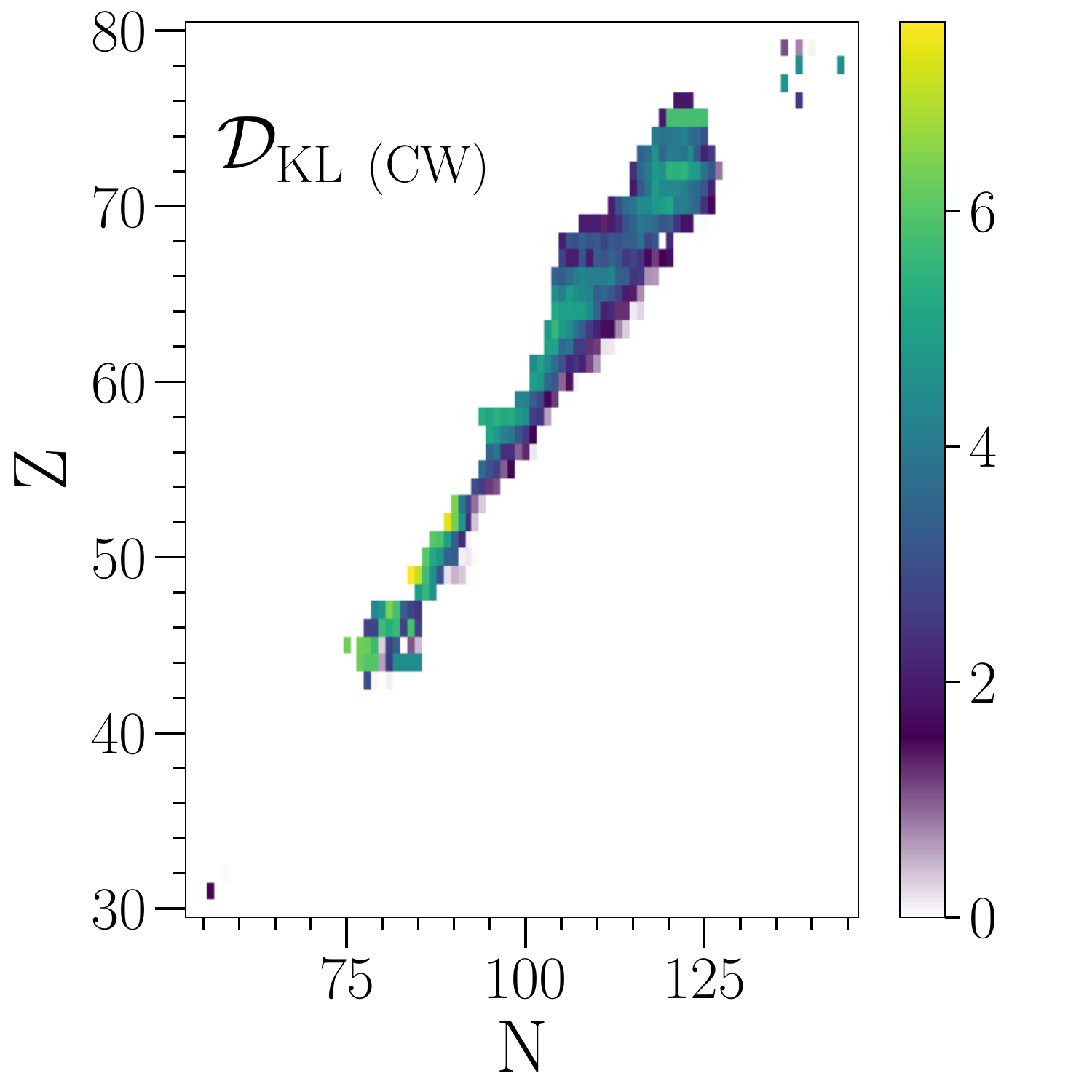}
\includegraphics[width=0.23\textwidth]{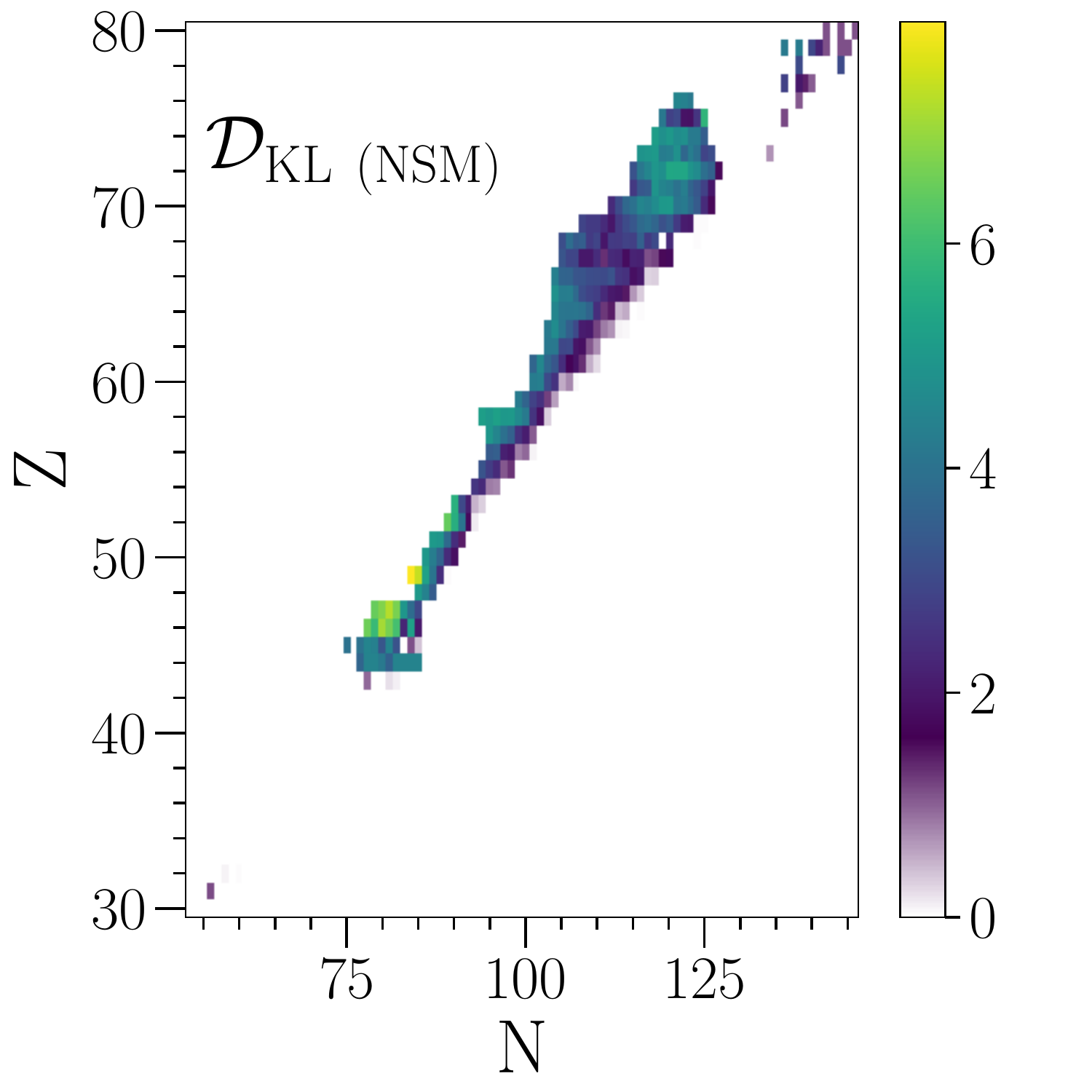}
\caption{KL divergence for low entropy hot wind (top left). KL
  divergence for high entropy hot wind (top right). KL divergence for
  cold wind (bottom left). KL divergence for neutron star mergers
  (bottom right). Nuclei in yellow represent the greatest information
  gain, while less informative nuclei are colored light
  purple.\label{fig:dir}}
\end{figure}

The theoretical information for each r-process scenario, using only
the nuclei which are both measurable in FRIB and important in
determining the r-process abundances, is summarized in
Fig.~\ref{fig:logth}. From the point of view of the r-process models
from Ref.~\cite{Mumpower16}, the nuclei with the largest $F$ values
and the smallest theoretical information are $^{134}\mathrm{Cd}$,
$^{136}\mathrm{Sn}$, $^{130}\mathrm{Pd}$, and $^{130}\mathrm{Pd}$ for
the LEHW, HEHW, CW, and NSM scenarios, respectively. 
%Fig.~\ref{fig:logpost} shows the $\mathrm{log}({\sigma_{\mathrm{post}}})$
%for all four astrophysical processes. This figure is consistent with
%Figs. ~\ref{fig:masses} and ~\ref{fig:logth} as combining the latter
%two figures will never decrease the information gain that appears in
%Fig. ~\ref{fig:logpost}. 
The information gain is summarized in
Fig.~\ref{fig:dir}. FRIB can measure nuclei near stability with a
smaller uncertainty, so the information gain is not maximized for the
nuclei above with the smallest theoretical information but for nearby
nuclei closer to stability. The mass measurements which optimize the
information gain are $^{133}\mathrm{In}$, $^{136}\mathrm{Sn}$,
$^{133}\mathrm{In}$, and $^{133}\mathrm{In}$ for the LEHW, HEHW, CW,
and NSM scenarios, respectively.

The values of $F$ from Ref.~\cite{Mumpower16} have an arbitrary
normalization, so we cannot quantitatively compare the different
r-process scenarios with each other. Our formalism only allows one to
determine the most informative mass measurements within one of the
four scenarios. Nevertheless, it appears that some nuclei, like
$^{133}\mathrm{In}$ and $^{134}\mathrm{In}$ are important masses to
measure independent of which r-process scenario is closer to reality.
%Although nuclei cannot quantitatively be compared to each other
%across each table (i.e.$^{133}\mathrm{In}$ having different values of
%$D_{\mathrm{KL}}$), the frequency and relative maximum
%$D_{\mathrm{KL}}$ should be compared. Based on tables II-IV,
%$^{133}\mathrm{In}$ would be the most beneficial isotope to study
%relative to r-process abundances.

For $D_{\mathrm{KL}}$ relative to r-process abundances, the
theoretical standard deviation has more of an impact than the
experimental standard deviation. This is in contrast to
$D_{\mathrm{KL}}$ relative to theoretical nuclear mass models, where
the theoretical standard deviation had very little impact in
$D_{\mathrm{KL}}$. The origin of this effect is that the F-values
found in Ref. ~\cite{Mumpower16} have a much larger range of four
orders of magnitude, ($0.01$ to $101.39$), as compared to the range of
uncertainties in the theoretical nuclear mass models from the previous
section (only a factor four). Since the F values cover a wider range,
they have more of an ability to impact the final value of
$D_{\mathrm{KL}}$. This can be seen by comparing the the top right
panel of Fig.~\ref{fig:masses} and each panel of Fig.~\ref{fig:logth}
with each panel in Fig.~\ref{fig:dir}. Fig.~\ref{fig:logth} and
Fig.~\ref{fig:dir} have the same general shape, while
$\sigma_{\mathrm{ex}}$ from Fig.~\ref{fig:masses} has some, but less
of an impact on $D_{\mathrm{KL}}$. This follows for the most part,
unless $\sigma_{\mathrm{ex}}$ is very small compared to
$\sigma_{\mathrm{th}}$, as in $^{134}\mathrm{Cd}$. $^{134}\mathrm{Cd}$
appears in tables ~\ref{tab:lehwtab} and ~\ref{tab:cwtab} with a
signifacntly larger $\sigma_{\mathrm{th}}$ than any other nucleus, but
is overshadowed by the small $\sigma_{\mathrm{ex}}$ of
$^{133}\mathrm{In}$, so there is a balance between these two values.

While this work was being completed, the $^{133}{\rm In}$ mass was
measured~\cite{Izzo21}. The next most informative measurement depends
on the most likely r-process scenario, though clearly the
($Z=50$,$N=82$) region of the periodic table is a source of many
informative mass measurements.

\section{Discussion and Conclusion}
For FRIB to make the most out of their resources, $D_{\mathrm{KL}}$
should be as large as possible. This means that for $D_{\mathrm{KL}}$
relative to theoretical mass models, the experimental information is
large compared to the theoretical information. In contrast, if the
theoretical information is large, regardless of the size of the
experimental information, $D_{\mathrm{KL}}$ will be small due to
having a large uncertainty in the theoretical models. If
$\sigma_{\mathrm{ex}}$ is small, regardless of the size of the
theoretical information, $D_{\mathrm{KL}}$ will be large, simply due
to the fact that there is a lot of information to gain. The least
uncertainty and therefore most information gain will occur in the
latter scenario, where $\sigma_{\mathrm{ex}}$ is small.

For $D_{\mathrm{KL}}$ relative to r-process abundances, the
theoretical information has more of an impact than the experimental.
This is due to having a larger range of theory values for the
r-process abundances as compared to the theoretical mass models. Both
$\sigma_{\mathrm{ex}}$ and $\sigma_{\mathrm{th}}$ do have an impact on
$D_{\mathrm{KL}}$, and the nuclei with the largest values of
$D_{\mathrm{KL}}$ in all four astrophysical scenarios occur when both
$\sigma_{\mathrm{ex}}$ and $\sigma_{\mathrm{th}}$ are small.

Our method for computing the KL divergence relative to the theoretical
mass models presumes that the new mass measurement will be equal to
the value predicted by theory. We have averaged over the theoretical
mass models in order to attempt to suppress our dependence on the
systematics of any one mass formula. However, if it is determined that
many of the mass predictions suffer from the same systematic flaw,
then this will impact the result.

In the future, facilities could utilize this function to their
advantage. Ideally, after each nucleus at a given facility is studied,
it is added to the experimental model, and this calculation is
performed again. This would allow for the most accurate measurements
of $D_{\mathrm{KL}}$ given the most up to date experimental
measurements. Using this model, facilities worldwide could collaborate
and nuclear physics could be studied at the most efficient rate
possible.

\begin{acknowledgments}
  JNF and AWS were supported by NSF grants PHY 1554876. ZM was
  supported in part by the U.S. Department of Energy under Grants No.
  DE-FG02-88ER40387, No. DE-NA0003909, and No. DE-SC0019042. ZM also
  benefited from support by the National Science Foundation under
  Grant No. PHY-1430152 (JINA Center for the Evolution of the
  Elements). AWS was additionally supported by NSF grant PHY 2116686
  and the U.S. DOE Office of Nuclear Physics.
\end{acknowledgments} 

\bibliographystyle{apsrev}
\bibliography{all}

\end{document}